\documentstyle[aps,prl,floats,epsf]{revtex}
\tighten
\draft

\begin{document}

\twocolumn[\hsize\textwidth\columnwidth\hsize\csname
@twocolumnfalse\endcsname
\title{Shell Model Description of Isotope Shifts in Calcium  \\}
\author { E. Caurier$^1$, K. Langanke$^2$, G. Mart\'{\i}nez-Pinedo$^{2,3}$ \\
F. Nowacki$^4$, and P. Vogel$^5$}
\address{$^1$ Institut de Recherches Subatomiques, Universit\'{e}
Louis Pasteur, F-67037 Strasbourg, France \\
$^2$ Institute of Physics and Astronomy, University of Aarhus,
DK-8000 Aarhus C, Denmark \\
$^3$ Department f\"{u}r Physik und Astronomie, Universit\"{a}t Basel,
CH-4056 Basel, Switzerland \\
$^4$ Laboratorie de Physique Th\'{e}orique, Universit\'{e}
Louis Pasteur, F-67037 Strasbourg, France \\
$^5$Department of Physics, California Institute of Technology,
         Pasadena, CA 91125, USA \\
}
\date{\today}
\maketitle

\begin{abstract}
Isotope shifts in the nuclear charge radius of even and odd 
calcium isotopes are calculated within the nuclear shell model.
The model space includes all configurations of nucleons in the 
$2s, 1d_{3/2}, 1f_{7/2}, {\rm and} ~2p_{3/2}$ 
orbits. The shell model describes well the energies of the intruder 
states in  Sc and Ca,
as well as the energies of the low-lying $2^+$ and $3^-$
states in the even Ca isotopes. The characteristic features of
the isotope shifts, the parabolic dependence on $A$ and the
prominent odd-even staggering, are well reproduced by the model.
These features are related to the partial breakdown of 
the $Z = 20$ shell closure caused by promotion, due to the
neutron-proton interaction, of the $ds$ shell protons into the
$fp$ shell.
\end{abstract}


]
\narrowtext

{\it Introduction~~}
The appearance of shell gaps associated with magic nucleon numbers is
one of the cornerstones in nuclear structure. However, it has become
increasingly evident in recent years that these magic numbers,
and the corresponding shell closures, might get eroded with
increasing neutron excess. To understand the origin of this erosion and
to identify whether this trend applies to all magic numbers is one of
the great challenges in nuclear structure with,
among other things, strong astrophysical
implications. A prominent example is the magic neutron number $N=20$
which vanishes in proton-deficient nuclei with 
$Z\leq12$~\cite{Motob} . This erosion
of the shell closure has been related to cross-shell proton-neutron
interaction which correlates the $2s$ and $1d_{3/2}$ proton orbitals
with the $1f_{7/2}$ and $2p_{3/2}$ neutron orbitals and leads to
appreciable deformation~\cite{War}. If cross-shell correlations are indeed the
mechanism for the shell erosion, then first indications are already
visible in the stable calcium isotopes, as we will argue in this Letter. Our
argument is based on the understanding and explanation of the nuclear
charge radii, $\langle r_c^2 \rangle$, in the calcium isotopes which 
by itself have been a challenge for nuclear
theory for the last two decades.

Experiments based on optical isotope shifts as well
as on muonic atom data (see Ref. \cite{Fricke} for a summary and original 
references, in particular \cite{Palmer}), 
revealed a characteristic parabolic shape of the 
isotope shifts with a pronounced odd-even staggering
when neutrons fill the $f_{7/2}$ orbit and the mass
number changes from $A = 40$ to $A = 48$. 
Thus, the challenge consists of understanding ($i$) why $^{48}$Ca is
not larger than $^{40}$Ca, $(ii)$ why the maximum of 
$\langle r_c^2 \rangle$ is reached in $^{44}$Ca, $(iii)$ why odd
$A$ calcium isotopes are considerably smaller than the average of their 
even neighbors, and $(iv)$ last but not least, the actual magnitude 
of the isotope shifts.

There have been numerous attempts to explain these findings.
If they are indeed related to the cross-shell proton-neutron
correlations, one expects that in order to describe them,
one must include higher than 2p-2h correlations.
This is necessary in order to reproduce the odd-even
staggering and, at the same time, shell effects
related to the neutron shell closures at $N$ = 20
and 28. This conjecture is supported by the fact that
mean field calculations, which usually aim at describing
nuclear masses, deformation parameters, and radii over a large
region of nuclear masses and charges, cannot account for
the details of the calcium isotope shifts. The dependence
of $\langle r_c^2 \rangle$ on $A$ is usually featureless.
Some of the approaches, however, are able to account at least for
the property $(i)$, i.e., the near equality of $\langle r_c^2 \rangle$
in $^{40}$Ca and $^{48}$Ca. To this category belong
calculations based on the Hartree-Fock method with Skyrme
interactions \cite{Werner}, relativistic mean field methods
\cite{Lalazissis}, and the extended Thomas-Fermi  model with the
Strutinsky-integral \cite{Aboussir}.

An extension of the mean field method with pairing, the so-called
local energy-density functional approach \cite{Fayans}, 
can describe most of the features mentioned above by introducing
three- and four-body forces via the density dependence of pairing,
and adjusting the corresponding free parameters.
 
That three-body forces can do the job has been recognized
long time ago. In  \cite{Caurier80} the isotope shifts in Ca
were quite well described using the configuration mixing within
the $(f_{7/2}d_{3/2})^n$ shell model space, and introducing 
an isoscalar particle-hole force whose strength 
was proportional to the number of valence
nucleons. The origin of that dependence and the magnitude
of the corresponding coupling constant, however, remained
unexplained.

Below we show that the modern shell model can describe
the isotope shifts in calcium consistently
and without the need for many-body forces or
adjustable parameters. The basic mechanism of
the $A$ dependence of the charge radius $\langle r_c^2 \rangle$
remains the same, however, as in Ref. \cite{Caurier80}.
Due to the configuration mixing across the $Z = 20$ shell
boundary, protons are lifted from the $sd$ to the $fp$
shell, resulting in the increase of $\langle r_c^2 \rangle$
equal to
\begin{equation}
\delta r_c^2 (A) = \frac{1}{Z} n_{fp}^{\pi} (A) \cdot b^2 ~,
\end{equation}
where $Z = 20$, $b$ is the oscillator parameter which we 
assume remains constant for $A = 40 - 48$, and the
number of protons in the $fp$ shell, $n_{fp}^{\pi}(A)$,
is the calculated quantity. Below we show how  it is
is calculated and how one can relate it to other
manifestations of breaching of the $Z = 20$ shell
boundary.

 
{\it Shell model~~}
Progress in the application of modern nuclear shell model has
been facilitated by the development of numerical codes 
(the $m$-code ANTOINE and the $J$-coupled code NATHAN
\cite{codes}) that
can be used with relatively large valence spaces.
The selection of the appropriate single-particle space,
and the corresponding effective interaction
(in particular the choice of its uncertain
monopole part \cite{DZ}) is the most important
starting point of any  shell model application.

Since here we are interested in the description of calcium
isotopes, it is imperative to include states in the
vicinity of the $N = Z = 20$ shell boundary. Therefore, the chosen
valence space consists of the $d_{3/2}, s_{1/2}, f_{7/2}$
and $p_{3/2}$ subshells for both protons and neutrons.
(Thus $^{28}$Si represents the inert core.) This
valence space, first used in Ref.\cite{hasper}, has the advantage
that the existing codes make it possible to describe
all Ca isotopes without truncation.
(The largest dimension, 34,274,564 in the $m$-scheme basis, is encountered
for the ground state of $^{43}$Ca.) The other advantage
is the essential absence of the spurious center-of-mass 
motion (no attempt to suppress it has been made).

For this valence space we start with the 
two-body matrix elements (TBME) of Ref. \cite{n28}, which  
are defined with respect to the $^{16}$O core. The
single particle energies are now modified to reproduce
the $^{29}$Si spectrum.
The interaction of Ref. \cite{n28}
was built in  blocks ($sd-sd$, $sd-fp$ and $fp-fp$), 
which incorporate
the higher-order excitations (core polarization, $2p2h$ excitations etc...). 
In particular, the $2p2h$ effects consist mainly 
in pairing renormalizations. As we will include explicitly such mixing, 
and to avoid double counting, a schematic pairing
hamiltonian was subtracted from the interaction.
Finally, the cross $sd-fp$ monopoles of Ref. \cite{n28} 
were adjusted to  the masses of neutron-rich isotopic chains.
They were here retuned to reproduce the gap at $^{40}$Ca and 
the  spectra of $^{39}$K and $^{41}$Ca.
   
\vspace{0.1cm}

\begin{figure}[h]
\epsfysize=2.9in \epsfbox{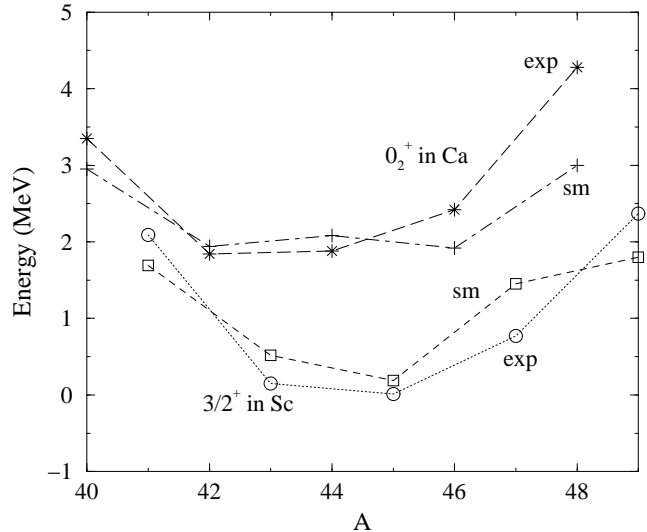}
\caption{\protect{\leftskip=3pc\rightskip=3pc\noindent
Excitation energies of intruder states in Sc and in even Ca
isotopes. The experimental (circles) and calculated
(squares) energies of the $3/2^+$ states in Sc and the 
experimental (stars) and calculated (crosses) 
energies of the excited $0_2^+$ states
in Ca are shown.}}
\label{fig:intr}
\end{figure}

{\it Results~~}
As indicated in Eq. (1) above, we assume that the main
cause of variation of the charge radius with changing
neutron number is the lifting of protons from the
$sd$ shell into the larger $fp$ shell. This feature
then suggests that there should be a correlation
between the isotope shifts and other manifestations
of the incomplete $Z = 20$ shell closure, e.g.
the appearance of low-lying intruder states.
In Fig. \ref{fig:intr} the experimental 
excitation energies of the
intruder $3/2^+$ states in odd-$A$ Sc isotopes
are compared with the calculated ones. The agreement
is very satisfactory, and in particular the lowering
of these states in $^{43}$Sc and $^{45}$Sc is properly
described.

The excited $0_2^+$ states in the even Ca isotopes also
belong to the intruder category. Their excitation
energies are shown in Fig. \ref{fig:intr} as well.
Again, the agreement between the experimental and 
calculated energies is quite good. (The larger discrepancy
in $^{48}$Ca is presumably caused by the absence of
the neutron $f_{5/2}$ subshell in the model space.)
These  $0_2^+$ states are dominantly $4p-4h$ states
involving both neutron and proton excitations.
The parabolic behaviour of both intruder excitations
in Fig. \ref{fig:intr} suggests that the configuration
mixing between the naive shell model configurations
and the intruders involving proton excitations from
the $sd$ shell will likely be most pronounced near $A = 44$
where the corresponding states are lowest.

The occupation numbers for both protons and
neutrons, $n_{fp} (A)$, in the ground states of
all $A = 40 - 48$ Ca isotopes were then evaluated
in the full shell model space. It is illustrative
to consider, however, how these numbers change
when more and more nucleons are allowed to be
excited from the naive configuration 
$(ds)^{6\pi + 6\nu},(fp)^{(A-40)\nu}$.
Varying the  number of nucleons, 
$t$, allowed
to be excited from the naive configuration above,
from $t = 2$
to its maximum value $t = A - 28$ we find
that the number of holes in the $ds$ shell
saturates very slowly. 
For example, in $^{42}$Ca and $^{43}$Ca one
has to allow $t$ = 6 and 7, respectively, 
just to reach the halfway point of  
the final $n_{fp} (A)$,
and $t$ = 10 and 11 to achieve saturation.
This means that the ``lifting'' is a complicated
process, involving substantial rearrangement
of many nucleons. This can be seen also in
the fact that approximately equal numbers of
protons and neutrons are lifted from the $ds$ shell.
It is therefore difficult to identify a simple cause,
or a definite component of the hamiltonian, as
the driving force of this effect.

In the full space the quantity $n_{fp} (A)$ 
is 1.10 in $^{40}$Ca,
reaches its maximum of 1.97 in $^{44}$Ca, and then 
decreases again to 0.78 in $^{48}$Ca. 

\begin{figure}[h]
\epsfysize=2.9in \epsfbox{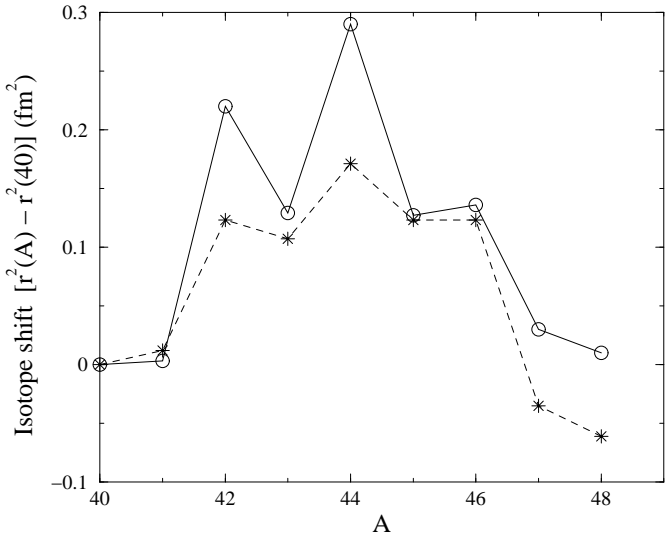}
\caption{\protect{\leftskip=3pc\rightskip=3pc\noindent
Isotope shifts in calcium. The experimental data (circles
connected by a solid line) and the shell model results, Eq.(1),
(stars connected by a dashed line) are shown.}}
\label{fig:is}
\end{figure}

Knowing the quantities $n_{fp}$
one can calculate the isotope
shifts from Eq. (1) by subtracting the corresponding
$\delta r_c^2 (A = 40)$ from $\delta r_c^2 (A)$.
(Nucleon form factors contribute negligibly
to the isotope shift and thus are neglected.)
The results, calculated with a constant
$b = 1.974$ fm as in \cite{Caurier80},
are shown in Fig. \ref{fig:is} and compared
with the experimental values. The trends, i.e. the
properties $(i)$ - $(iii)$ are clearly well reproduced,
but the magnitude of the calculated shifts is smaller
than the experiment suggests.

This cannot be attributed
to our choice of the oscillator parameter $b$. Neither it
can be cured by replacing the harmonic
oscillator wave functions by the Woods-Saxon single-particle
wave functions,
which changes the isotope shifts shown in Fig. 
\ref{fig:is} only very little, provided the
core $\langle r^2 \rangle$ is assumed to be
independent of $A$.
(Although assuming a very slight increase
of $b(A)$, i.e. a gradual increase
of the core  $\langle r^2 \rangle$
from $A = 40$ to $A = 48$, might 
``straighten'' the curve a bit.) Moreover, choosing
the recommended average dependence of $b$ on the
mass number $A$ and isospin $T$ \cite{Zucker99}
($b^2 = 1.07 A^{1/3} (1 - (2T/A)^2) exp(3.5/A)$ fm$^2$)
would make the radius of $^{48}$Ca considerably larger
than the radius of $^{40}$Ca. The $b(A,T)$  dependence,
which on average follows  \cite{Zucker99},
clearly is subject to local shell effects.
Most likely, the calculated amplitude of the isotope shift
might be modified if the so far neglected effects of the
filled $d_{5/2}$ and empty $f_{5/2}$ orbits are added.


{\it Collective states~~}
When considering isotope shifts, one also has to take into account
the effect of the zero-point motion associated with the
surface vibrational modes (see \cite{Bertsch} for a systematic
approach to this question. There it is stressed
that only the low-lying low-multipolarity surface
vibrational states contribute significantly). 
The collective states increase
the mean square radius by
\begin{equation}
\delta r^2 = \frac{R_0^2}{4 \pi} \sum_{\lambda} \beta_{\lambda}^2 ~,
\label{eq:coll}
\end{equation}
where $R_0$ is the equivalent sharp nuclear radius, and $\beta_{\lambda}$
is the vibrational amplitude of the mode $\lambda$,
related to the corresponding $B(E\lambda,0 \rightarrow J=\lambda)$
value. What matters for the isotope shifts, naturally, is not
the absolute value of the amplitudes $\beta_{\lambda}$, but their
variation with $A$.

\begin{figure}[h]
\epsfysize=2.9in \epsfbox{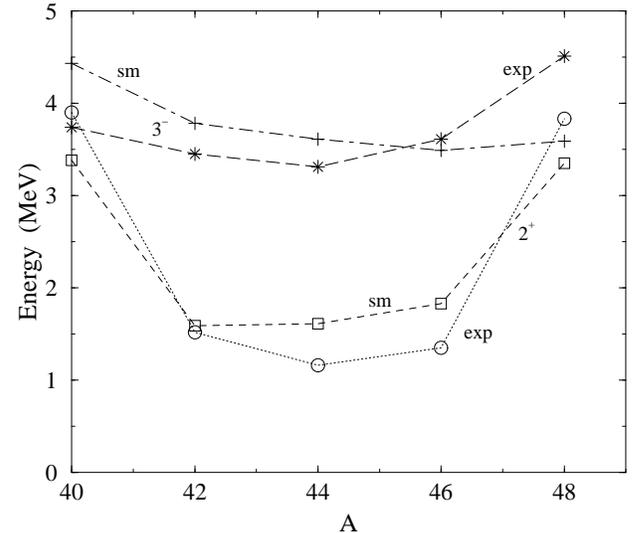}
\caption{\protect{\leftskip=3pc\rightskip=3pc\noindent
Excitation energies of the $2^+$ and $3^-$ states
in the even Ca isotopes.}}
\label{fig:coll}
\end{figure}

\begin{figure}[h]
\epsfysize=2.9in \epsfbox{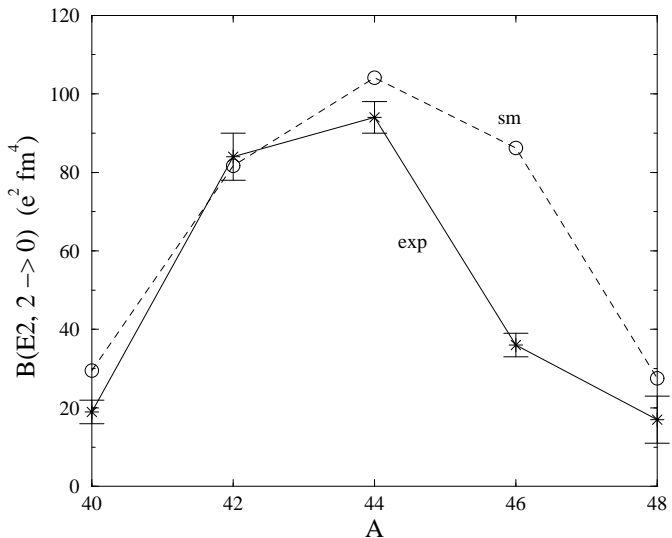}
\caption{\protect{\leftskip=3pc\rightskip=3pc\noindent
Comparison of the experimental and shell
model $B(E2, 2 \rightarrow 0)$ values
in the even Ca isotopes.}}
\label{fig:be2}
\end{figure}

Formula (\ref{eq:coll}) was derived as a correction to the
mean square radius calculated using the mean-field methods. 
On the other hand, since the shell model includes all
correlations of nucleons in the valence shells, blind use
of it would amount to at least partial double counting.

The energies of the lowest lying $2^+$ and $3^-$ states are compared 
to the shell model in Fig. \ref{fig:coll}. The good agreement
there shows not only that the chosen interaction describes this
important spectroscopic data well, but that the zero-point
motion effect on the nuclear radius is (at least most of it)
automatically included.

In addition, in Fig.~\ref{fig:be2} we compare the 
experimental and shell model $B(E2, 2 \rightarrow 0)$
values. (The calculations were performed with the
usual effective charges $e_p = 1.5, e_n = 0.5$.)
Again, the agreement is quite good, strengthening our 
belief that most of the effect of the collective states
is already contained in our calculation.


{\it Discussion~~}
The present calculation shows that
the $Z = 20$ and $N = 20$ shell boundaries are clearly 
not absolute, and that substantial configuration mixing
involving nucleons from the $ds$ shell is present even in the
Ca isotopes. 

This finding is in line with the previously recognized
and perhaps more dramatic ``islands of inversion'' related
to the configuration mixing involving the $N = 20$
shell boundary in neutron rich nuclei (see Ref. 
\cite{Caurier98,Caurier00}). Here we are encountering 
a similar situation, now however in stable nuclei near
or at $N = Z$.

Is there any other evidence for the incomplete shell closure
in $^{40}$Ca? Indeed, the study of the $(n,p)$ and $(p,n)$
reactions \cite{Park,Cagt,Tadd}
on  $^{40}$Ca revealed that the integrated Gamow-Teller 
strength $B(GT)$ = $1.6 \pm 0.1$ below 15 MeV. Obviously,
if the $ds$ shell in $^{40}$Ca were really closed,
the $B(GT)$ would vanish.
Note, that without
the full inclusion of the $f_{5/2}$ and $d_{5/2}$ orbit we can
only estimate the total Gamow-Teller strength in $^{40}$Ca.  
Without these orbits we obtain the total $B(GT)$=0.53 using the usual
quenching of 0.744 \cite{quench}. 
When we allow the $t = 1$ (i.e. one particle or
one hole) excitations involving the $f_{5/2}$ and $d_{5/2}$ orbits,
we obtain $B(GT)$ = 2.75 with quenching, in fair agreement
with the experiment. Clearly, some of the strength
associated with these orbits will be above 15 MeV,
and should not be counted.

In conclusion,  the present calculation 
shows that the isotope shifts, and the
position of the intruder states in Ca isotopes, can be well
described by large scale shell model calculations.
The judicious choice of the valence space and the
corresponding effective hamiltonian is the key ingredient
to this success. Thus, the challenge to the nuclear structure
theory, described in the introduction to this work, has
been largely met. No new forces or parameters are needed
to describe the dependence of the nuclear radius 
in the odd and even Ca isotopes on the mass number $A$.


{\it Acknowledgment~~} This work has been partially supported
by a grant of the Danish Research Council,
Carlsberg Foundation, Sweizerische Nationalfonds, and by the
U. S. Department of Energy. F. N. and P. V. thank the
Institute of Nuclear Theory, Seattle, where part of the work
was done, for hospitality. In addition, E. C. and P. V.
thank the Institute of Physics and Astronomy at the University
of Aarhus for its hospitality.


\end{document}